\documentstyle[prb,aps,preprint]{revtex}
\begin{document}
\topmargin-0.5cm
\addtolength{\textheight}{5mm}
\setcounter{page}{1}
\tighten
\widetext
\title{\mbox{}\vspace{-2cm}\\
{\normalsize\makebox[0mm][l]{}\hfill 
\hfill 
\makebox[0mm][r]{}}\vspace{5mm}\\
Laterally coupled few-electron quantum dots}
\author{Andreas Wensauer, Oliver Steffens, Michael Suhrke, and Ulrich R\"ossler}
\address{Institut f\"ur Theoretische Physik, 
Universit\"at Regensburg, D-93040 Regensburg, Germany}
\date{Version \today}
\maketitle
\begin{abstract}
We present ground-state calculations for laterally coupled quantum
dots containing 2, 4, and 8 electrons. As our emphasis is on spin
effects our results are obtained by applying spin-density functional
theory (SDFT). By varying the distance between the centers of the
coupled quantum dots, the transition from weak to strong coupling
situation is realized. For the 2-electron system we also apply the
Heitler-London approximation and analytical concepts to check the
reliability of SDFT calculations in this case. In addition we discuss
the features of the Coulomb staircase of laterally coupled quantum
dots in the weak and strong
coupling regimes in comparison to that of a circular parabolic
quantum dot.
\mbox{}

\end{abstract}

\narrowtext

\section{Introduction}
Semiconductor quantum dots are manmade nanoscale structures in which
electrons are confined in all three spatial directions similar to
the physical situation in atoms. As they show typical atomic
properties like discrete energy levels and shell structures they are
often referred to as artificial atoms\cite{kastner93}. However, in contrast to
natural atoms, in quantum dots the number of electrons $N$ is
tunable and the characteristic lengths of the system corresponding to
external confinement potential, electron-electron interaction,
and an applied magnetic field are of comparable size. Therefore,
these systems are ideal objects to study interaction effects such as transitions 
in the
ground-state spin configuration as a
function of the magnetic field. 

Starting from quantum dots as a structure which is well understood by now, more 
complex 
systems are conceivable and likely to have perspective in future applications 
\cite{zanardi99}$^,$\cite{imamura95}$^,$\cite{burkard99}. A simple example is 
analogous 
to a two-atom molecule consisting of two coupled quantum dots. 
The more principal properties of coupled quantum dots follow from
the simple model of e.g.\ two disk-like dots side by side in a plane
or on top of each other (lateral or vertical quantum-dot molecules
(QDMs)), respectively, which are the subject 
of experimental\cite{austing98}$^,$\cite{waugh95a}$^,$\cite{waugh95b} and
theoretical\cite{imamura99}$^,$\cite{rontani99}$^,$\cite{yannouleas99}$^,
$\cite{nagaraja99}
studies. Tunneling and overlap of orbital wavefunctions lead to
bonding and antibonding states with defined parity, which are
delocalized over the two-dot system as in natural systems. The
advantage of QDMs is the tunability of the electron number and of the
coupling strength by proper design. The latter allows to
investigate the transition between weak and strong coupling.

The increased interest in coupled quantum dots is indicated by a
growing number of papers on this topic: Austing {\em et
  al.}\cite{austing98} measured addition spectra of vertical QDM in
different coupling regimes. The characteristic features found in
their spectra have been confirmed by Imamura {\em et
  al.}\cite{imamura99} and Rontani {\em et al.}\cite{rontani99} applying exact 
diagonalization techniques. These
ground-state calculations were performed for up to 6 electrons as a
function of the interdot distance and an external magnetic
field. Coulomb blockade effects in lateral QDMs have been measured by
Waugh {\em et al.}\cite{waugh95a}$^,$\cite{waugh95b}, who found
a conductance pattern similar to that of single
dots for strong coupling, whereas the weakly coupled systems showed a pairing of
conductance peaks. Theoretical work on ground-state properties of
lateral QDMs was performed by Yannouleas and
Landman\cite{yannouleas99} 
using an unrestricted Hartree-Fock
approach and by Nagaraja {\em et al.}\cite{nagaraja99} applying SDFT.

In this paper we investigate the electronic structure of the ground
state of two laterally coupled (identical) quantum dots as a
function of the interdot distance. In our
calculations, applying SDFT, we focus on
spin-dependent effects in few-electron quantum
dot molecules containing 2, 4, and 8 electrons. 
In the case of 4 (8) electrons we find a change of spin configuration
with decreasing coupling strength from $S_z=1$ ($S_z=1$) to $S_z=0$ ($S_z=2$).
For 2 electrons SDFT yields a transition from spin
singlet to spin triplet. It turns out, however, that the latter
result is an artifact of the SDFT, as can be shown (at least for our
choice of the confinement potential) by using alternative approaches.
Furthermore, we show the characteristic Coulomb staircases for the chemical 
potential in the weak and strong coupling limits, respectively.

For all our calculations\cite{wensauer99} we assume that the two dots are 
coupled
in a quantum-mechanically coherent way. This means we use
delocalized electron states extended over the whole quantum dot
molecule whereas states localized to the left or right dot do not exist
in this description. Note, that this model does not contain
all aspects of the actual physical situation where an
electron occupying a covalent state interacts with other electrons
and charges in the surrounding semiconductor material. These
interactions may cause a dephasing of
the quantum mechanical wavefunction leading to a breakdown of the
delocalized state\cite{oosterkamp98}. Up to now no theory is available which 
allows a
reliable approximation for dephasing rates in realistic
nanostructures. 

The outline of this paper is the following: In the next section we
describe our model of the QDM  and discuss the
single-particle spectrum of the non-interacting system. In Sec.\ III
we sketch the basics of SDFT and of the numerical method for the
simultaneous solution of the Kohn-Sham (KS) equations and
calculation of the
selfconsistent potentials. The following part is devoted to the
2-electron QDM. We compare the SDFT results with
exact (analytical) results and with the Heitler-London approach to
point out a particular problem of SDFT. Sec.\ V relates the spin
configuration for 4 and 8 electrons to the single-particle
spectrum and summarizes the corresponding SDFT results. Finally we
show the typical Coulomb staircase of the addition energies for the
weak and the strong coupling limits, respectively.

\section{The model and its single-particle system}
We consider quantum dots prepared by laterally confining
two-dimensional electrons in a GaAs/AlGaAs heterostructure (material
parameters of GaAs: $m^\ast=0.067 m_e$, $\varepsilon=12.4$). As the
lateral confinement is much weaker than that in growth direction ($z$-direction) 
we adopt the standard
situation of the electronic quantum limit and describe the electron
density by $n(x,y,z)=n(x,y)\delta (z)$. Thus, our system is defined by the
effective two-dimensional Hamiltonian
\begin{eqnarray}
\label{hamiltonian}
H&=&T+V+W\nonumber\\
&=&\sum_{j=1}^N\left(\frac{{\bf p}_j^2}{2m^\ast}+
V({\bf r}_j)\right)+\frac{e^2}{4\pi\varepsilon\varepsilon_0}
\sum_{j=1}^N\sum_{k=j+1}^N\frac{1}{\left|{\bf r}_j-{\bf r}_k\right|}
\end{eqnarray}  
with ${\bf p}_j$ and ${\bf r}_j$ being vectors in the
$xy$-plane. Our choice of the external confinement potential $V({\bf r})$
is 
\begin{equation}
V({\bf r})=\frac{1}{2}m^\ast\omega_0^2
\min\left(\left({\bf r}-{\bf L}\right)^2,
\left({\bf r}+{\bf L}\right)^2\right), 
\end{equation}
where the coupling strength depends on the distance $d=\left|2 {\bf L}\right|$ 
between the potential minima at $\pm {\bf L}$ in the $xy$-plane
(Fig.\ 1). Along the line between these minima there is a barrier
of height
\begin{equation}
V_{\rm B}(d)=\frac{1}{2}m^\ast\omega_0^2\left(\frac{d}{2}\right)^2.
\end{equation}
This model potential describes a single parabolic quantum
dot for $d=0$ as the extreme case of strong coupling and two 
separate quantum dots of the same shape in the
weak-coupling limit for
$d\rightarrow\infty$. Due to the broken axial symmetry of
$V({\bf r})$ for $d\neq 0$ the numerical requirements are increased
in comparison with those for a single circular quantum dot.

In Fig.\ \ref{qdmnww} we show the lowest energy levels of the noninteracting
electrons in dependence on the distance between the dot centers for
the typical confinement energy $\hbar\omega_0=3$ ${\rm meV}$. For
$d=0$, when the confinement potential of the QDM degenerates to the simple
parabolic potential with axial symmetry,
the spectrum shows the typical shell
structure of an isotropic harmonic oscillator. For large distances
$d$ we obtain the same spectrum, now twofold because the system
consists of two identical but completely separated quantum dots. By
switching on the coupling (or lowering $d$ starting from the
weak-coupling limit) we recover the anticipated properties of a
diatomic molecule: the energies decrease (increase) with respect to
the reference level due to formation of bonding (antibonding)
states, which have even (odd) parity as indicated by solid (dashed)
lines. The intermediate coupling regime is
dominated by crossings leading to the
rearrangement of the level structure. In the weak-coupling limit the
nth bundle of the twofold level structure with energy $E\approx n\hbar\omega_0$
consists of $n$ bonding states and $n$ antibonding states.  

\section{SDFT and KS equations}
In order to include the electron-electron interaction we
employ the spin-density functional theory (SDFT). It is the
generalization of the DFT formalism, originally established by
Hohenberg, Kohn, and Sham\cite{hohenberg64}$^,$\cite{kohn65}, to 
spin-polarized systems\cite{barth72} by including the
coupling of the magnetization to an applied magnetic field. Accordingly,
the Hohenberg-Kohn (HK) theorem has to be modified with respect
to the spin degrees of freedom\cite{barth72}. For this case, it states that two
 different
nondegenerate ground-state wavefunctions $|\Psi\rangle$ and
$|\Psi^\prime\rangle$ always yield different tupels $(n({\bf r}),{\bf
  m}({\bf r}))\neq (n^\prime({\bf r}),{\bf m}^\prime({\bf r}))$ of
electron density $n({\bf r})$ and magnetization ${\bf m}({\bf
  r})$. This is sufficient to establish a functional of the total
energy with the usual functional properties
\begin{equation}
E_{V_0,{\bf B}_0}[n,{\bf m}]=F_{\rm HK}[n,{\bf m}]+\int{\rm d}{\bf
  r}\left[V({\bf r})n({\bf r})-{\bf B}({\bf r})\cdot{\bf m}({\bf r})\right]
\end{equation}
and the universal HK functional 
\begin{equation}
F_{\rm HK}[n,{\bf m}]=\langle\Psi[n,{\bf m}]|T+W|\Psi[n,{\bf
  m}]\rangle .
\end{equation}
In the limit ${\bf B}\rightarrow 0$ considered here, the SDFT scheme can
yield a spin-polarized
ground state for even electron numbers due to Hund's 
rule\cite{koskinen97}$^,$\cite{steffens98a}. These
effects have already attracted much interest and will also be discussed
in this paper.

The spin-degree of freedom is considered in the Kohn-Sham (KS) 
equations\cite{barth72} by
assuming the total spin $S_z$ in $z$-direction to be a good quantum number
\begin{eqnarray}
{\Bigg\lbrace} -\frac{\hbar^2}{2m^\ast}\nabla^2+V({\bf r})+
\frac{e^2}{4\pi\varepsilon\varepsilon_0} \int{\rm d}{\bf r}^\prime
\frac{n({\bf r}^\prime)}{\left|{\bf r}-{\bf r}^\prime\right|}\nonumber\\+
V_{\rm XC}^\sigma([n_+,n_-],{\bf r}]{\Bigg\rbrace}\varphi_j^\sigma({\bf r})
=\varepsilon_j^\sigma\varphi_j^\sigma({\bf r})
\end{eqnarray}
with the spin $\sigma=\pm$ in $z$-direction and the KS energies
$\varepsilon_1^\sigma\le\varepsilon_2^\sigma\le...$ . For a
system containing $N$ particles we calculate the occupation numbers
of the KS levels in the ground state due to
\begin{eqnarray}
\gamma_j^\sigma=1 & \hspace*{1cm} & \varepsilon_j<\mu\\
0\le\gamma_j^\sigma\le 1 & \hspace*{1cm} & \varepsilon_j=\mu\\ 
\gamma_j^\sigma=0 & \hspace*{1cm} & \varepsilon_j>\mu
\end{eqnarray}
($\mu$: chemical potential) with the constraints 
\begin{eqnarray}
\sum_j \gamma_j^\sigma&=&N^\sigma\\
N^++N^-&=&N.
\end{eqnarray}
This leads to the spin densities
\begin{equation}
n_\sigma({\bf r})=
\sum_j\gamma_j^\sigma\left|\varphi_j^\sigma({\bf r})\right|^2,
\end{equation}
the total density
\begin{equation}
n({\bf r})=n_+({\bf r})+n_-({\bf r}),
\end{equation}
and the magnetization in $z$-direction ($\mu_B$: Bohr's magneton)
\begin{equation}
m_z({\bf r})=-\mu_{\rm B}\left(n_+({\bf r})-n_-({\bf r})\right).
\end{equation}
The exchange-correlation (xc) potentials 
\begin{equation}
V_{\rm XC}^\sigma([n_+,n_-],{\bf r})=
\frac{\delta E_{\rm XC}[n_+,n_-]}{\delta n_\sigma({\bf r})}
\end{equation}
are defined as functional
derivatives of the xc energy functional
\begin{eqnarray}
E_{\rm XC}[n_+,n_-]&=&F_{\rm HK}[n_+,n_-]-
\frac{1}{2}\frac{e^2}{4\pi\varepsilon\varepsilon_0}
\int{\rm d}{\bf r}\int{\rm d}{\bf r}^\prime
\frac{n({\bf r})n({\bf r}^\prime)}
{\left|{\bf r}-{\bf r}^\prime\right|}\nonumber\\
&&-T_{\rm S}[n_+,n_-].
\end{eqnarray}
($T_{\rm S}[n_+,n_-]$ denotes the kinetic energy functional of the KS
system.)
The total ground-state energy $E_0$ of the interacting system can be calculated
from 
\begin{eqnarray}
E_0&=&\sum_{j,\sigma}\gamma_j^\sigma\varepsilon_j^\sigma-
\frac{1}{2}\frac{e^2}{4\pi\varepsilon\varepsilon_0}
\int{\rm d}{\bf r}\int{\rm d}{\bf r}^\prime
\frac{n({\bf r})n({\bf r}^\prime)}
{\left|{\bf r}-{\bf r}^\prime\right|}\nonumber\\
&&+E_{\rm XC}[n_+,n_-]-
\sum_\sigma\int{\rm d}{\bf r}
\,V_{\rm XC}^\sigma([n_+,n_-],{\bf r})n_\sigma({\bf r}).
\end{eqnarray}
Concerning the xc potentials we apply the local spin density
approximation (LSDA) 
\begin{equation}
E_{\rm XC}[n_+,n_-]\approx
\int{\rm d}{\bf r}\left(n_+({\bf r})+n_-({\bf r})\right)
\varepsilon_{\rm XC}(n_+({\bf r}),n_-({\bf r}))
\end{equation}
and the Pad\'e approximation of 
$\varepsilon_{\rm XC}(n_+({\bf r}),n_-({\bf r}))$ in two dimensions
following Tanatar and Ceperley\cite{tanatar89}.

For the numerical solution of the KS equations we calculate all
quantities with spatial dependence on a two-dimensional grid in real
space. The damped gradient iteration\cite{blum92} ensures a
simultaneous solution of the KS equations together with the corresponding
self-consistent potentials. This method uses the iteration scheme
\begin{equation}
|\Psi^{(k+1)}\rangle=H^{-1}|\Psi^{(k)}\rangle
\label{idea}
\end{equation}
with the KS Hamiltonian $H$ (only positive eigenvalues),
which converges to the ground-state wavefunction. The inversion of the 
Hamiltonian is performed approximately\cite{blum92}$^,$\cite{reinhard99} 
using
\begin{eqnarray}
\label{asympt}
|\Psi^{(k+1)}\rangle &\approx&{\Bigg \lbrace}{\Bigg \lbrack} 1-{\tilde x_0}
\left(\frac{p_x^2}{2m^\ast}+V_x^\infty+E_{x0}\right)^{-1}\cdot\nonumber\\
&&\cdot\left(\frac{p_y^2}{2m^\ast}+ V_y^\infty+E_{y0}\right)^{-1}\cdot
\nonumber\\
&&\cdot\left(H-\langle\Psi^{(k)}|H|\Psi^{(k)}\rangle\right){\Bigg \rbrack}
{\Bigg \rbrace}
|\Psi^{(k)}\rangle
\end{eqnarray}
instead of Eq.\ (\ref{idea}),
where $\tilde x_0$, $E_{x0}$, $E_{y0}$ are iteration parameters and 
$V_x^\infty$, $V_y^\infty$ the asymptotic contributions of
the external potential for $|{\bf r}|\rightarrow\infty$.
Excited state KS wavefunctions are calculated using the same iteration scheme 
with an additional
orthogonalization routine.
The Hartree and xc potentials depending on the densities of the KS 
wavefunctions are recalculated 
in each step. Thus, the KS Hamiltonian is modified in each step of the 
iteration. 

\section{Results for the two-electron QDM}
Fig.\ \ref{ksqdm2} depicts the KS energies for a quantum
dot molecule containing two electrons as a function of the distance
$d$ between the centers of the molecule (we assume $\hbar\omega_0=3$
${\rm meV}$).

Starting with $d=0$ the QDM degenerates into a
circular parabolic quantum dot showing the known shell structure of
the noninteracting case. Due
to the closed shell for 2 electrons the ground state is not
spin-polarized. This ground-state spin configuration is stable for
increasing $d$ as long as the energy gap between the levels (a) and
(b) in the single-particle spectrum (Fig.\ \ref{qdmnww}) is not too
small. At
$d\approx 4.5$ ${\rm a}_0^\ast$ (${\rm a}_0^\ast=9.79$ ${\rm nm}$ denotes the
effective Bohr's radius for GaAs) the unpolarized ground state changes
into a polarized one. This state with aligned spins persists as SDFT
ground state even for $d\rightarrow\infty$, because the energy gap
between states (a) and (b) continues to decrease with increasing
$d$. This result, similar to that obtained e.g. by Nagaraja {\em et
  al.}\cite{nagaraja99}, differs from those of the physically
analogous problems of the vertical QDMs\cite{austing98}$^,$\cite{bryant93} and 
from the
hydrogen molecule whose ground states are spin singlets.

The SDFT result is also in contrast to a mathematical theorem 
stating that the spin
configuration of the ground state of a two-electron quantum dot
molecule, with our definition of the external confinement potential,
should be a spin singlet state: As the total Hamiltonian $H$, Eq.\ 
(\ref{hamiltonian}),
does not depend on spin coordinates we can focus on the spatial
wavefunctions. Due to Theorem XIII.47 of Ref. \onlinecite{reed78} the (spatial) 
ground
state wavefunction of the considered two-body-problem is positive and
nondegenerate, this means it is symmetric in the spatial coordinates. As a 
consequence of
Pauli's principle the ground state has to be a spin singlet state.

In view of this rigorous statement, the SDFT results of Fig.\ \ref{ksqdm2} for 
$d\ge 4.5$ ${\rm a}_0^\ast$ cannot be considered as true ground
state values (the same holds for the corresponding SDFT
densities). 

We test this finding by using a perturbation approach analogous to the
Heitler-London model for the hydrogen molecule, which in contrast to
SDFT has the advantage of yielding an upper limit for the ground
state energy. Similar calculations for other shapes of the external potential 
were done in Ref.\ \onlinecite{burkard99}.

For the two-electron QDM we rewrite the total
Hamiltonian (\ref{hamiltonian}) as
\begin{eqnarray}
H&=&\frac{{\bf p}_1^2}{2m^\ast}+\frac{1}{2}m^\ast\omega_0^2
\min\left(\left({\bf r}_1-{\bf L}\right)^2,
\left({\bf r}_1+{\bf L}\right)^2\right)\nonumber\\
&&\frac{{\bf p}_2^2}{2m^\ast}+\frac{1}{2}m^\ast\omega_0^2
\min\left(\left({\bf r}_2-{\bf L}\right)^2,
\left({\bf r}_2+{\bf L}\right)^2\right)\nonumber\\
&&+\frac{e^2}{4\pi\varepsilon\varepsilon_0}
\frac{1}{\left|{\bf r}_1-{\bf r}_2\right|}\nonumber\\
&=&H_1^{\rm R}+H_2^{\rm L}+U_1^{\rm R}+U_2^{\rm L}+W
\end{eqnarray}
with
\begin{eqnarray}
H_j^{\rm R/L}&=&\frac{{\bf p}_j^2}{2m^\ast}
+\frac{1}{2}m^\ast\omega_0^2\left({\bf r}_j\mp{\bf L}\right)^2\\
U_j^{\rm R/L}&=&\frac{1}{2}m^\ast\omega_0^2
\min\left(0,\pm4{\bf r}_j\cdot{\bf L}\right)\\
W&=&\frac{e^2}{4\pi\varepsilon\varepsilon_0}
\frac{1}{\left|{\bf r}_1-{\bf r}_2\right|}.
\end{eqnarray}
The ground-state wavefunctions of the shifted harmonic oscillators
$H^{\rm R/L}$ are denoted by $|\Psi^{\rm R/L}\rangle$.
So we can use the typical Heitler-London ansatz for the
(unnormalized) singlet $|\Psi^+\rangle$ and triplet
$|\Psi^-\rangle$ spatial wavefunctions
\begin{equation}
|\Psi^\pm\rangle=\frac{1}{\sqrt{2}}
\left(|\Psi^{\rm R}\rangle^{(1)}|\Psi^{\rm L}\rangle^{(2)}
\pm |\Psi^{\rm R}\rangle^{(2)}|\Psi^{\rm L}\rangle^{(1)}\right).
\end{equation}
to calculate the expectation value of the ground-state energy of the
system
\begin{eqnarray}
E^\pm_{\rm HL}&=&\frac{\langle\Psi^\pm|H|\Psi^\pm\rangle}
{\langle\Psi^\pm|\Psi^\pm\rangle}=\nonumber\\
&=&2\hbar\omega_0+\frac{\langle\Psi^{\rm R}|U^{\rm R}|\Psi^{\rm R}\rangle
+\langle\Psi^{\rm L}|U^{\rm L}|\Psi^{\rm L}\rangle}
{1\pm{\rm e}^{-2L^2/l_h^2}}\nonumber\\
&&+\frac{{^{(1)}\langle}\Psi^{\rm R}|{^{(2)}\langle}\Psi^{\rm L}|W|
\Psi^{\rm R}\rangle^{(1)}|\Psi^{\rm L}\rangle^{(2)}}
{1\pm{\rm e}^{-2L^2/l_h^2}}\nonumber\\
&&\pm\frac{{\rm e}^{-L^2/l_h^2}\left(\langle\Psi^{\rm R}|U^{\rm R}
|\Psi^{\rm R}\rangle
+\langle\Psi^{\rm L}|U^{\rm L}|\Psi^{\rm L}\rangle\right)}
{1\pm{\rm e}^{-2L^2/l_h^2}}\nonumber\\
&&\pm\frac{{^{(1)}\langle}\Psi^{\rm R}|{^{(2)}\langle}\Psi^{\rm L}|W|
\Psi^{\rm L}\rangle^{(1)}|\Psi^{\rm R}\rangle^{(2)}}
{1\pm{\rm e}^{-2L^2/l_h^2}}.
\end{eqnarray}
Note that all matrix elements are real quantities.
$l_h=\sqrt{\hbar/m^\ast\omega_0}$ denotes the characteristic oscillator length.
The result for the matrix elements is
\begin{eqnarray}
&&\langle\Psi^{\rm R}|U^{\rm R}|\Psi^{\rm R}\rangle
+\langle\Psi^{\rm L}|U^{\rm L}|\Psi^{\rm L}\rangle=\nonumber\\
&&=2\hbar\omega_0
\left(-\frac{L}{l_h\sqrt{\pi}}{\rm e}^{-L^2/l_h^2}
+\frac{L^2}{l_h^2}\left(1-{\rm erf}\left(L/l_h\right)\right)\right)
\end{eqnarray}
\begin{eqnarray}
&&{^{(1)}\langle}\Psi^{\rm R}|{^{(2)}\langle}\Psi^{\rm L}|W|
\Psi^{\rm R}\rangle^{(1)}|\Psi^{\rm L}\rangle^{(2)}=\nonumber\\
&&\hspace*{3cm}=\hbar\omega_0\sqrt{\frac{\pi}{2}}\frac{l_h}{{\rm a}_0^\ast}
{\rm e}^{-L^2/l_h^2}{\rm I}_0(L^2/l_h^2)
\end{eqnarray}
\begin{eqnarray}
\langle\Psi^{\rm R}|U^{\rm R}|\Psi^{\rm L}\rangle
+\langle\Psi^{\rm L}|U^{\rm L}|\Psi^{\rm R}\rangle
=-2\hbar\omega_0\frac{L}{l_h\sqrt{\pi}}{\rm e}^{-L^2/l_h^2}
\end{eqnarray}
\begin{eqnarray}
{^{(1)}\langle}\Psi^{\rm R}|{^{(2)}\langle}\Psi^{\rm L}|W|
\Psi^{\rm L}\rangle^{(1)}|\Psi^{\rm R}\rangle^{(2)}
=\hbar\omega_0\sqrt{\frac{\pi}{2}}\frac{l_h}{{\rm a}_0^\ast}
{\rm e}^{-2L^2/l_h^2}
\end{eqnarray}
with the modified Bessel's function ${\rm I}_0(x)$ and the error
function ${\rm erf}\left(x\right)$ \cite{gradsteyn80}.

In Fig.\ \ref{heitler} we compare the ground-state energies resulting from SDFT
calculations with the Heitler-London approach. 
For $0\le d\le 4.5$ ${\rm a}_0^\ast$ the SDFT singlet ground-state
energies are lower than the HL energies. As the comparison between
SDFT and exact diagonalization for 2-electron quantum dots shows
good agreement\cite{steffens99}, we expect this to hold also for the QDM with
small $d$. Moreover, the HL triplet energy diverges for
$d\rightarrow 0$ and becomes larger than the HL singlet
energy. Therefore, we regard the SDFT results for small $d$ as a
better approximation of the ground state than the HL
results\cite{fussnote}.  

If, however, $d$ is increased beyond $4.5$ ${\rm a}_0^\ast$ the HL
energies become lower than the SDFT results and replace them as
ground-state energies. It can be shown analytically that for
sufficiently large $d$ the HL energies fulfill the relation
$E^+_{\rm HL}<E^-_{\rm HL}$. Thus, in the limit $d\rightarrow\infty$ the HL 
results are
consistent with the mathematical theorem quoted above.
In addition, in the weak-coupling limit the interaction is reduced to the 
classical
Coulomb repulsion between point charges and the total energy approaches 
asymptotically  
$E=2\hbar\omega_0+e^2/(4\pi\varepsilon\varepsilon_0 d)$. This is also consistent 
with the 
energies of the HL approach.

The energy difference of $0.36$ ${\rm meV}$ between $E^-_{\rm SDFT}$
and $E^+_{\rm HL}$ in the weak-coupling limit can be ascribed to
the self-interaction within the SDFT scheme: The charge density 
of a weakly coupled quantum dot molecule is mainly localized in its centers 
(charge 1 $e^-$ per center). However, within the SDFT
formalism the energy of each charge in its own Coulomb field is
counted additionally leading to an enhanced total energy\cite{fussnote2}.  
To illustrate this, let us to apply the SDFT concept to a
single-particle problem. For a circular parabolic quantum dot with 
$\hbar\omega_0=3$ ${\rm meV}$ this gives a ground-state energy of $3.18$ 
${\rm meV}$, i.e.\ the self-interaction error is $0.18$ ${\rm
  meV}$. This value has to be doubled for the $d\rightarrow\infty$
limit of the QDM. Thus, we identify the difference $E^-_{\rm
  SDFT}-E^\pm_{\rm HL}$ as the self-interaction energy in this limit. 

The energy difference between the singlet state and the
triplet state in the SDFT approach can be qualitatively explained as an xc 
effect:
Both spin configurations show similar total electron densities. Consequently
the contributions from the Hartree energy and the spin-independent
xc energy can be taken to be approximately equal for singlet
and triplet states. However, the triplet total energy is in
addition lowered by spin-dependent xc effects leading to the energy difference
in Fig.\ \ref{heitler}.  

To summarize this section: for small $d$ ($0\le
d\le 4.5$ ${\rm a}_0^\ast$) we find the SDFT (singlet) energies to be the
better approximation to the ground state than the HL results, while
for larger $d$ ($d\ge 4.5$ ${\rm a}_0^\ast$) the SDFT energies
become too large due to the self-interaction error and the
HL results provide the better approximation of the ground-state
energy. 

\section{Results for QDMs containing 4 and 8 electrons}
In this section we present our results for the ground-state energy and 
spin-configuration for artificial molecules containing 4 and 8 electrons 
as a function of the interdot distance $d$. Our calculations show that a 
spin-polarized 
ground state is accompanied by a (quasi-)degeneracy at the Fermi level in the 
noninteracting single-particle spectrum whereas unpolarized ground states 
require a sufficiently
large gap  between the highest occupied and lowest unoccupied single-particle 
levels of the noninteracting spectrum.

Single dots with 4 electrons have a spin-polarized ground state
($S_z=1$) due to Hund's
rule\cite{koskinen97}$^,$\cite{steffens98a}. This is the case also
for the QDM with $d=0$ (Fig.\ 5). The spin-polarized ground state
persists with increasing $d$ as long as the xc energy
overcompensates the energy costs of putting electrons with parallel
spins into the bonding and
antibonding states ((b) and (c) in Fig.\ \ref{qdmnww}) evolving
from the second level at $d=0$. For $d>2$ ${\rm a}_0^\ast$ it
becomes energetically more favorable to occupy both
the bonding level (a) and antibonding level (b) (Fig.\ \ref{qdmnww}) with two 
particles
forming an $S_z=0$ ground state. This configuration remains stable
even in the weak coupling limit and reflects the noninteracting
particle picture (Fig.\ \ref{qdmnww}) which does not show level
crossings at the Fermi energy for $N=4$ and large $d$. This result
is in agreement with the work of
Nagaraja {\em et al.}\cite{nagaraja99} (laterally coupled dots), Rontani 
{\em et al.}\cite{rontani99}, and
Imamura {\em et al.}\cite{imamura99} (vertically coupled dots). On
the other hand, Nagaraja {\em et al.}\cite{nagaraja99} do not find a
spin-polarized ground state in the strong coupling limit, because
their confinement potential obtained by simulating external gates does
not approach that of a circular parabolic dot
as in our case.

Although even in the limit of large $d$ a quantum-mechanical
coupling (delocalized bonding and antibonding states) between the
dots can be observed within the SDFT scheme the resulting density
(Fig.\ \ref{qdm4_10n}) and energy (Fig.\ \ref{qdm4_e}) tend towards a 
semiclassical
picture: the densities of the dots are well separated and the total
energy of the two coupled dots consists of twice the (quantum-mechanical)
energy of a single dot containing 2 electrons and the energy of classical
repulsion of two point-like charges (2$e^-$) in the distance $d$ (see inset of 
Fig.\ \ref{qdm4_e}). In contrast to the 2-electron QDM, this test for the 
asymptotics of the total energy being fulfilled underlines the reliabilty of the
 SDFT results
for 4 electrons.

Turning to QDMs with 8 electrons (Fig.\ \ref{ksqdm8}) we find for
$d=0$ the expected spin-polarized ground state
($S_z=1$)\cite{koskinen97}$^,$\cite{steffens98a}. Increasing $d$ up to
$7.5$ ${\rm a}_0^\ast$ this state persists with double-occupancy of
the levels (a), (b), and (c) of Fig.\ \ref{qdmnww} and
single-occupancy of the quasi-degenerate levels (d) and (e) with
aligned spins. At $d\approx 7.5$ ${\rm a}_0^\ast$ the levels (c),
(d), (e), and (f) of Fig.\ \ref{qdmnww} get close enough that
an alignment of all four spins at the Fermi level can lower the
total energy. Therefore, we find a spin-polarized ground state in the
weak coupling regime ($d>8$ ${\rm a}_0^\ast$) in agreement with
Nagaraja {\em et al.}\cite{nagaraja99}. Although the
spin-aligned ground state was ruled out in the case of 2 electrons
we believe that the $S_z=2$ ground state for the 8-electron artificial
molecule in the weak-coupling regime is reliable. This is emphasized by the
correct asymptotics of the total energy in the limit $d\rightarrow\infty$
(see inset of Fig.\ \ref{qdm8_e}). Exact
diagonalizations\cite{imamura99}$^,$\cite{rontani99} for vertically
coupled dots performed for up to 6 electrons yield a
spin-polarized ground state ($S_z=1$) for the 6-electron artificial
molecule in the weak coupling regime, too, indicating that spin-polarized
ground states can exist in principle for QDMs. 

Also for $N=8$, in the weak-coupling limit, the total energy and
density can be understood in a semiclassical picture: the densities
(for sufficiently large $d$) are well separated (Fig.\ \ref{qdm8_10n}) and
the total energy follows that of two identical dots each containing
4 electrons plus the Coulomb repulsion between 
two point-like charges (4$e^-$) in distance $d$. However, besides the
quantum mechanical phenomenon of delocalized bonding and antibonding
states we find a (nonclassical) xc energy related effect leading to the
spin-polarized ground state.   

Yannouleas and Landman\cite{yannouleas99} find from unrestricted Hartree-Fock
calculations electron localization in the form of electron puddles and Wigner
supermolecules for a ratio $\lambda$ between the oscillator length and Bohr's
radius of about $1.4$. In contrast, (exact) Quantum Monte Carlo 
calculations\cite{egger99} indicate a transition from the Fermi liquid to the 
Wigner crystal regime if $\lambda$ is increased beyond a critical value of 4. 
Thus, our calculations performed for $\lambda\approx 2$ are obviously in the
Fermi liquid regime.  

\section{Chemical potential}
Although recent experiments succeeded in identifying molecular
properties of coupled quantum dots like bonding and antibonding
states\cite{oosterkamp98} the most convenient comparison between theoretical and
experimental results is based on the chemical potential of the
system. The chemical potential is defined as difference of total energies
$\mu(N)=E(N)-E(N-1)$ which can be measured by capacitance
spectroscopy\cite{waugh95a}$^,$\cite{waugh95b}$^,$\cite{tarucha96}. 

As it turns out, the dependence of $N$ on $\mu$ (Coulomb staircase)
exhibits features which depend on the coupling
strength\cite{waugh95a}$^,$\cite{waugh95b}$^,$\cite{nagaraja99}.
While in the strong coupling limit the capacitance spectrum of the QDM
is similar to that of the single quantum dot, it exhibits 
small spacings between the chemical
potentials for $N=1, 2$, $N=3, 4$, $N=5, 6$ etc.\ for the weakly
coupled QDM which correspond to the
pairing of peaks observed in capacitance
spectroscopy\cite{waugh95a}$^,$\cite{waugh95b}. Explanations for
this phenomenon provided in the literature\cite{waugh95a} are based on
states localized at the centers of the molecule and neglect quantum mechanical 
coupling. Using these assumptions we
can easily estimate the dependence of $\mu$ on $N$ for our system:
If the QDM contains an even
number $N$ of electrons one may assume that $N/2$ electrons are
localized at each dot. The total energy of the artificial molecule
$E_{\rm M}(N)$
can be estimated by the energy of the single dots containing $N/2$
electrons ($E_{\rm D}(N/2)$) and the Coulomb repulsion between the
dots which is approximated as Coulomb energy between point-like
charges
\begin{equation}
E_{\rm M}(N)=2E_{\rm D}\!\left(\frac{N}{2}\right)+
\frac{e^2(N/2)^2}{4\pi\varepsilon\varepsilon_0 d}
\mbox{, $N$ even.}
\end{equation}
For odd $N$ one assumes $(N+1)/2$ electrons to be in one dot and
$(N-1)/2$ in the other one and the approximation for the total energy
yields 
\begin{eqnarray}
E_{\rm M}(N)&=&E_{\rm D}\!\left(\frac{N+1}{2}\right)+E_{\rm D}\!
\left(\frac{N-1}{2}\right)\nonumber\\
&&+\frac{e^2(N+1)(N-1)/4}{4\pi\varepsilon\varepsilon_0 d}
\mbox{, $N$ odd.}
\end{eqnarray}
Using these equations we can calculate the chemical potential
$\mu_{\rm M}(N)=E_{\rm M}(M)-E_{\rm M}(N-1)$ of the QDM for even and odd $N$
\begin{equation}
\mu_{\rm M}(N)=E_{\rm D}\!\left(\frac{N}{2}\right)-E_{\rm D}\!\left(\frac{N}{2}
-1\right)+
\frac{e^2N/2}{4\pi\varepsilon\varepsilon_0 d}
\mbox{, $N$ even,}
\end{equation}
\begin{equation}
\mu_{\rm M}(N)=E_{\rm D}\!\left(\frac{N+1}{2}\right)-E_{\rm D}\!\left(
\frac{N-1}{2}\right)+
\frac{e^2(N-1)/2}{4\pi\varepsilon\varepsilon_0 d}
\mbox{, $N$ odd.}
\end{equation}
Consequently, the addition energies which are necessary to add a
further electron are 
\begin{equation}
\Delta\mu_{\rm M}(N)=\frac{e^2}{4\pi\varepsilon\varepsilon_0
  d}\mbox{, $N$ even,}
\end{equation}
\begin{equation}
\Delta\mu_{\rm M}(N)=\mu_{\rm D}\!\left(\frac{N+1}{2}\right)-
\mu_{\rm D}\!\left(\frac{N-1}{2}\right)\mbox{, $N\ge3$ odd.}
\end{equation}
These equations indicate that in a weakly coupled quantum dot with
strongly localized wavefunctions the addition of an electron to a
configuration with an even electron number is energetically more expensive as it 
corresponds to an addition of an electron to a single quantum dot. On
the other hand, adding an electron to an odd number of particles is
relatively cheap because the addition energy only depends on the
interdot Coulomb
 repulsion which is relatively small for large $d$ and gives
rise to the typical pairing of conductance peaks. With growing
electron numbers this model fails as the increasing Coulomb repulsion
caused by more extended charge distributions is not included. This
effect destroys the splitting of the paired peaks. 

A conceptional difficulty of this
model is that it is based on localized states yielding an
asymmetric charge distribution for odd electron numbers. In our SDFT
calculations, however, 
the delocalized states also lead to the pairing
of the conductance peaks in the weak-coupling regime (Fig.\
\ref{coulstair}) while providing symmetric
electron densities for all $N$. Therefore, we would like to
emphasize, that the pairing of peaks does not depend on
electron states localized at one center of the QDM, but can be
caused by delocalized wavefunctions as well. In addition, delocalized
states provide the advantage that it is not necessary to determine a
transition from a coherent to an incoherent regime when the distance
$d$ between the centers is continuously increased.

\section{Conclusions}
We have studied the ground-state properties of QDMs
containing 2, 4, and 8 electrons as a function of the distance $d$
between the centers of the molecule using SDFT. This concept
includes spin effects in addition to the classical Coulomb
repulsion. The results obtained 
for 4 and 8 electrons are reliable over the whole range of $d$ and
can be predicted from the single-particle spectrum of the
noninteracting system for not too weak confinement. For 4 electrons
we find a transition from a spin-polarized to a spin-unpolarized
configuration with decreasing coupling strength. The 
ground state of the 8-electron QDM can be characterized by a
spin-polarized ground state (2 parallel spins) for strong coupling
and a spin-polarized configuration with 4 aligned spins in the
weak-coupling limit. In
contrast, the two-electron QDM is well described
by SDFT only in the strong-coupling regime whereas for large $d$ a HL
approach yields better results. Finally we have discussed the Coulomb
staircase diagram for the weak and the strong coupling regime and
identified their characteristic features considering the background
of delocalized states in an artificial molecule.       

This work was funded by Deutsche Forschungsgemeinschaft (Grants
No. SFB 348 and Ro 522/16).



\noindent
\unitlength1cm
\begin{figure}
\caption{Model potential for a QDM. In this
  picture the interdot distance is $d=13$ ${\rm a}_0^\ast$ and the
  strength of the parabolic potential near the centers is
  $\hbar\omega_0=3$ ${\rm meV}$.
\label{potential}
}
\end{figure}

\noindent
\begin{figure}
\unitlength1cm
\caption{Lowest levels in the spectrum of the
  noninteracting system as a function of the interdot distance
  $d$. The levels are classified as bonding and antibonding states
  with even and odd parity, respectively.
\label{qdmnww}
}
\end{figure}

\noindent
\begin{figure}
\unitlength1cm
\caption{Lowest KS levels for the 2-electron QDM as
  a function of the interdot distance $d$ (the 2 lowest levels
  are occupied). In the shaded area at $d\approx4.5$ ${\rm
    a}_0^\ast$ SDFT calculations yield a transition from a
  spin-unpolarized to a spin-polarized ground state. 
\label{ksqdm2}
}
\end{figure}

\noindent
\begin{figure}
\unitlength1cm
\caption{Comparison between the total energies of the SDFT
  calculation ($E_{\rm SDFT}^{\pm}$) and of the Heitler-London approach
  ($E_{\rm HL}^{\pm}$) for the 2-electron QDM. $+$ ($-$) denotes singlet 
  (triplet) energies.
\label{heitler}
}
\end{figure}

\noindent
\begin{figure}
\unitlength1cm
\caption{Lowest KS levels for the 4-electron QDM as
  a function of the interdot distance $d$ (the 4 lowest levels
  are occupied). In the shaded area at $d\approx 2$ ${\rm
    a}_0^\ast$ SDFT calculations yield a transition from a
  spin-polarized ($S_z=1$) to a spin-unpolarized ($S_z=0$) ground state. 
\label{ksqdm4}
Inset: Ground-state energy of the 4-electron QDM
  (crosses). The solid line describes the asymptotics for large
  $d$ in a quasi-classical picture as a combination of the ground
  state energy of two single dots containing 2 electrons ($E_{\rm
    Dot}^{(2)}=11.418$ ${\rm meV}$) and the Coulomb repulsion
  between two point-like charges (2 $e^-$) in distance $d$.
\label{qdm4_e}
}
\end{figure}

\noindent
\begin{figure}
\unitlength1cm
\caption{Ground-state density of a 4-electron QDM
  for $d=10$ ${\rm a}_0^\ast$. The character of two uncoupled dots is
  emphasized by the localization of the density in the centers of the
  molecule. 
\label{qdm4_10n}
}
\end{figure}

\noindent
\begin{figure}
\unitlength1cm
\caption{Lowest KS levels for the 8-electron QDM as
  a function of the interdot distance $d$ (the 8 lowest levels
  are occupied). In the shaded area at $d\approx7.5$ ${\rm
    a}_0^\ast$ SDFT calculations yield a transition from a
  spin-polarized ($S_z=1$) to a higher spin-polarized ($S_z=2$) ground state. 
\label{ksqdm8}
Inset: Ground-state energy of the 8-electron QDM
  (crosses). The solid line describes the asymptotics for large
  $d$ in a quasi-classical picture as a combination of the ground
  state energy of two single dots containing 4 electrons ($E_{\rm
    Dot}^{(4)}=41.064$ ${\rm meV}$) and the Coulomb repulsion
  between two point-like charges (4 $e^-$) in distance $d$.
\label{qdm8_e}
}
\end{figure}

\noindent
\begin{figure}
\unitlength1cm
\caption{Ground-state density of a 8-electron QDM
  for $d=10$ ${\rm a}_0^\ast$. The character of two uncoupled dots is
  emphasized by the localization of the density to the centers of the
  molecule. 
\label{qdm8_10n}
}
\end{figure}

\noindent
\begin{figure}
\unitlength1cm
\caption{Coulomb staircase (derived from SDFT results) for the 
  weak-coupling regime (a) ($d=10$
  ${\rm a_0^\ast}$), for the strong coupling regime (c) ($d=3$ ${\rm
    a_0^\ast}$), and for the circular parabolic quantum dot (c) ($d=0$
  ${\rm a_0^\ast}$). The shaded areas in picture (a) indicate the
  pairing of conductance peaks. For higher electron numbers the
  pairing splits as a consequence of increasing Coulomb repulsion
caused by more extended charge distributions.
\label{coulstair}
}
\end{figure}


\begin{references}
\bibitem{fussnote}
{However,
for $0.5$ ${\rm a}_0^\ast\le d\le 4.5$ ${\rm a}_0^\ast$ the situation is 
vice versa: The HL
triplet state is lower in energy than the HL singlet state as
a consequence of the weak external confinement potential in comparison
to the relatively strong external potential in the hydrogen
molecule.}
\bibitem{fussnote2}
{The effect of self-interaction in DFT calculations strongly depends on 
the number of particles. In
the case of small electron numbers (2 or 3 particles) it can cause deviations 
whereas for larger
numbers it can be neglected.}
\end{references}
\end{document}